\documentclass[twocolumn,prl]{revtex4}
\pdfoutput=1
\usepackage{graphicx}
\usepackage{color}

\newcommand{\Ket}[1]{\vert \, #1 \, \rangle}

\newcommand{\be}{\begin{equation}}
\newcommand{\ee}{\end{equation}}
\newcommand{\bea}{\begin{eqnarray}}
\newcommand{\eea}{\end{eqnarray}}

\renewcommand{\phi}{\varphi}
\renewcommand{\epsilon}{\varepsilon}
\renewcommand{\vec}[1]{{\bf #1}}

\begin{document}

\title{
Detection of spin injection into a double quantum dot:\\
Violation of magnetic-field-inversion symmetry of nuclear polarization instabilities
}

\author{Mark S. Rudner$^1$ and Emmanuel I. Rashba$^{1, 2}$}
\affiliation{
$^{(1)}$ 
 Department of Physics,
 Harvard University, 
 Cambridge, MA 02138\\
$^{(2)}$
 Department of Physics,
 Loughborough University,
 Leicestershire LE11 3TU, UK
}


\begin{abstract}

In mesoscopic systems with 
spin-orbit coupling, 
spin-injection into quantum dots at zero magnetic field 
is expected under a wide range of 
conditions. However, up to now, a viable approach for experimentally identifying such 
injection has been lacking.
We show that electron spin injection into a spin-blockaded double quantum dot is dramatically manifested in the breaking of magnetic-field-inversion symmetry of nuclear polarization instabilities.  
Over a wide range of parameters, the asymmetry between positive and negative instability fields is 
extremely sensitive 
to the injected electron spin polarization
and allows for the detection of even very weak spin injection.
This phenomenon may be used to 
investigate 
the 
mechanisms of spin
transport, 
and may 
hold 
implications for spin-based information processing.

\end{abstract}

\maketitle

Time reversal symmetry is a fundamental law of nature, which places strong constraints on the types of behaviors which can occur in physical systems.
Based on the Onsager concept of microscopic reversibility close to equilibrium, general transport coefficients exhibit particular symmetries under the reversal of direction of an applied magnetic field $\vec{B}$ \cite{TheBook}.
Such magnetic-field-inversion symmetry is robustly observed to high precision throughout a wide variety of experiments. Therefore it is quite remarkable to find examples of phenomena where this symmetry is violated.
Moreover, such asymmetries can provide information about 
deviations from equilibrium.

Over the past several years, excitement about the prospect of spin-based information processing 
has led many authors to consider a variety of mechanisms for injecting and manipulating electron spins
in nanoscale devices. 
Such works have shown that, through the spin-orbit interaction, significant spin 
injection can be 
produced even in the absence of an applied magnetic field\cite{Perel03, KiselevKim03, Ohe, Eto, Krich, Sablikov}. 
The only restrictions on spin injection are imposed by the action of time reversal symmetry together with unitarity, and they allow spin injection in systems with more than one outgoing channel\cite{KiselevKim03, ZhaiXu, Krich}.
However, as shown in Ref.\cite{Krich}, 
coupling to an environment which breaks unitarity 
allows spin injection even into a single outgoing channel. 
Experimentally, spin injection through quantum point contacts has also been reported \cite{Rokhinson,Debray}. 
Thus 
spin injection 
appears to be a generic phenomenon.

Although less extensively studied, analogous mechanisms should lead to spin
injection of electrons into quantum dots in systems with 
spin-orbit coupling, at $\vec{B}=0$.
For quantum dots coupled to source and drain electrodes, 
unitarity is broken due to the coupling to phonons and Fermi reservoirs, and
there is no fundamental reason to expect a vanishing spin injection probability. 

Although such spin injection is expected under a wide range of circumstances, until now a viable method for its detection 
has been missing. 
In this paper, we demonstrate that 
spin injection can be manifested in a dramatic violation of magnetic-field-inversion symmetry in dc transport through spin-blockaded double quantum dots. 
Experiments in this regime \cite{Ono04, Koppens05, Pfund07} have demonstrated a variety of interesting nonlinear phenomena such as bistabilities and hysteresis, which are associated with the coupled dynamics of electron and nuclear spins. 
In particular, $\bf B$-inversion asymmetry has apparently been recently observed by the Delft group \cite{FK}.
To date, the theoretical treatment of instabilities 
\cite{Rudner1, Inarrea07} was based on the assumption of completely unpolarized injected electron spins. 
Here we use an extended version of the model of Ref.\cite{Rudner1} to show that 
spin injection breaks the magnetic-field-inversion symmetry of the dynamical instabilities.
Furthermore, in the regime where hyperfine-
and non-hyperfine-mediated decay rates are comparable, the degree of asymmetry is an extremely sensitive function of the injected electron spin polarization. 

Why is magnetic-field-inversion symmetry violated for this system?
The direction of the dc current flowing through the double dot breaks the time-reversal symmetry, even for very weak currents.
Through spin-orbit coupling, this violation of time-reversal by the direction of the current is converted into spin injection 
into the dot.
While the observation of spin injection {\it per se} does not reveal its mechanism, by
varying electrostatic gates which control the transport
of electrons in the lead and/or barrier regions, the phenomenon can be used to investigate the nature of spin
transport in the system.

Note that, in the presence of a large Zeeman splitting, electron spins may be injected with a high degree of polarization\cite{Amasha, Ren10}. However, such magnetically-induced spin injection is symmetric in the magnetic field $\bf B$, and therefore {\it does not} lead to magnetic-field-inversion asymmetry. Here we focus on the field-independent part of spin-injection which persists down to zero magnetic field. For simplicity, we further assume that magnetically-induced spin-injection is weak over the range of relevant fields. To illustrate the spin-injection induced magnetic-field-inversion-symmetry-breaking phenomenon most clearly, we start from the simple model of spin-blockaded transport proposed in Ref.\cite{OnoScience} and employed to investigate nuclear spin polarization instabilities in Ref.\cite{Rudner1}.
Although the details of the results are model-dependent, we expect the phenomenon itself to persist more generally, 
when specific properties of realistic experimental setups are taken into account.

\begin{figure}[t]
\includegraphics[width=3.4in]{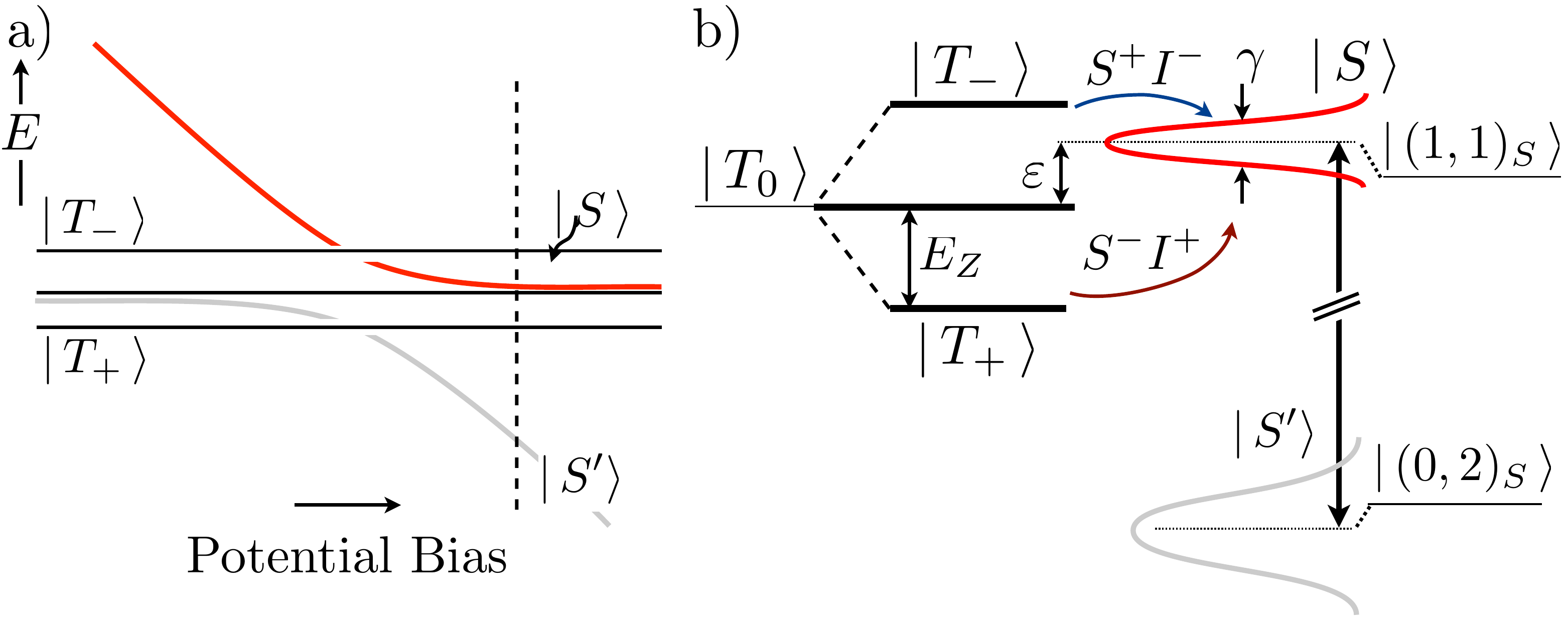}
  \caption[]{Energy level diagram of spin-blockaded double quantum dot.  
At large potential bias, 
hyperfine-assisted transitions between $\Ket{T_\pm}$ and the $(1,1)$ singlet state $\Ket{S}$ are accompanied by nuclear spin flips which lead to dynamical nuclear polarization and feedback through the Overhauser shift of the electron Zeeman energy $E_Z$, see text for details.
}
 \label{fig:Levels}
\end{figure}

The energy levels of the double quantum dot as a function of potential bias are depicted in Fig.\ref{fig:Levels}a. For large positive bias, indicated by the dashed vertical line, the ground electronic state $\Ket{S'}$ is a two-electron spin-singlet, with large weight in the ``(0,2)'' orbital configuration where both electrons occupy the right dot. In addition, the double dot supports ``(1,1)'' spin-singlet, $\Ket{S}$, and spin-triplet, $\Ket{T_{0,\pm}}$, states, in which electron density is nearly equally shared between the two dots. Because the triplet states are not directly coupled to the drain lead, the steady-state current in this spin-blockade regime is controlled by the rates of processes such as hyperfine exchange with nuclear spins which break 
electron spin conservation. 
These processes, in turn, can lead to dynamical polarization of nuclear spins.

We study nuclear polarization dynamics through a rate equation for the mean polarization $x= (N_+ - N_-)/(N_+ + N_-)$, where $N_+$ ($N_-$) is the population of nuclear spins oriented parallel (antiparallel) to the external magnetic field axis. For each electron that enters the dot, the probability of its decay resulting in a nuclear spin flip from down to up (up to down) is determined by the probability $P_\pm$ of having injected the state $\Ket{T_+}$ ($\Ket{T_-}$), and the relative competition between the hyperfine spin-exchange decay rate $W^{\rm HF}_\pm$ and the nuclear-spin-independent escape rate $W^{\rm in}$, which we take to be energy-independent. As a result, the nuclear polarization evolves according to 
\be
\label{xdot}
\dot{x} = \frac{I_0}{eN}\left(\frac{P_+ W_+^{\rm HF}}{W_+^{\rm HF} + W^{\rm in}} - \frac{P_- W_-^{\rm HF}}{W_-^{\rm HF} + W^{\rm in}}\right),
\ee
where $I_0$ is the total current through the device, $e$ is the electron charge, and $N = N_+ + N_-$ is the total number of nuclear spins in the system.
The injected electron spin polarization
is encoded in the factors $P_+$ and $P_-$.

We focus on the regime of large detuning, 
depicted in Fig.\ref{fig:Levels}b, where the energy of the singlet level $\Ket{S'}$ 
is far below the energies of the blockaded triplet states. 
Here the rates $W^{\rm HF}_\pm$ of nuclear spin flips arising from elastic hyperfine transitions between $\Ket{T_\pm}$ and $\Ket{S}$ are obtained from Fermi's Golden Rule:
\be\label{WHF}
W^{\rm HF}_\pm = \frac{2\pi}{\hbar}\frac{(1 \mp x)}{2} \mathcal{M}^2 f(\epsilon_\pm),\quad f(\epsilon) = \frac{\gamma/\pi}{\epsilon^2 + \gamma^2},
\ee
where $\epsilon_\pm = \epsilon \pm E_Z$ (see Fig.\ref{fig:Levels}b), 
and we assume a Lorentzian lineshape 
of width $\gamma$ for the decaying singlet 
state.
Here $E_Z = -\mu_e B + Ax$ is the effective Zeeman energy including the Overhauser shift $Ax$,
where $\mu_e = -g^*\mu_B$ is electron magnetic moment with $g^*$ the effective g-factor of the material ($g^* \approx -0.4$ in GaAs), $\mu_B$ is the Bohr magneton, $B = |\vec{B}|$ is the magnitude of the external magnetic field, and $A$ is the hyperfine coupling strength.
The matrix element $\mathcal{M} \sim A/\sqrt{N}$ for electron-nuclear spin exchange is set by the typical scale
of the random transverse hyperfine field. 
The factor $\frac12 (1 \mp x)$ counts the available phase space for finding a properly oriented (down or up) nuclear spin to flip.

We now seek the steady state values of nuclear polarization, obtained by setting $\dot{x} = 0$ in Eq.(\ref{xdot}). Transforming to the set of dimensionless parameters $\tilde\epsilon = \epsilon/A$, $\tilde B = \mu_eB/A$, $\tilde\gamma = \gamma/A$, $\tilde m = \mathcal{M}^2/(\hbar W^{\rm in}A)$, and the spin-injection coefficient $\eta = (P_+ - P_-)/(P_+ + P_-)$, the steady state values of the nuclear polarization are given by the third-order algebraic equation  
\be\label{Fx}
F(x) \equiv ax^3 + bx^2 + cx + d =0,
\ee
with 
\bea\label{abcd}
\nonumber a &=& 1,\quad b =  \eta(2\tilde\epsilon + \tilde\gamma \tilde m - 1) - 2\tilde B,\\
\nonumber c &=& \tilde\epsilon^2 + \tilde B^2 + \tilde\gamma^2 - 2\tilde\epsilon(1 + \eta\tilde B) + 2 \tilde B\eta,\\
d &=& 2\tilde B\tilde \epsilon - \eta(\tilde B^2 + \tilde\epsilon^2 + \tilde\gamma^2 + \tilde\gamma\tilde m).
\eea
Typically, 
$A \approx 100\ \mu$eV, while the singlet-triplet splitting $\epsilon$ and 
level width $\gamma$ can be on the $\mu$eV scale or less. Therefore, below we 
take $\tilde\epsilon, \tilde\gamma \ll 1$. 
As mentioned above, 
we 
disregard Zeeman-splitting-induced spin injection, 
which would produce 
an effect 
even $B$. Thus we consider $\eta$ as {\it field independent}.

A cubic equation with real coefficients, such as that in Eq.(\ref{Fx}), may have either one or three real solutions, depending on the values of the coefficients. Each such solution, which corresponds to a steady state of Eq.(\ref{xdot}), can be stable or unstable, depending on whether the flow $\dot{x}$ tends to restore or amplify small deviations from the steady state. In parameter regimes where Eq.(\ref{xdot}) possesses two stable fixed points $\dot{x} = 0$, the system is {\it bistable} and will typically exhibit hysteresis and/or possible switching\cite{switching}. 
As a parameter such as the magnetic field $B$ is varied, bistability disappears at {\it bifurcation points}, where two real roots of Eq.(\ref{Fx}) annihilate and become a complex-conjugate pair.
 
\begin{figure}[t]
\includegraphics[width=3.4in]{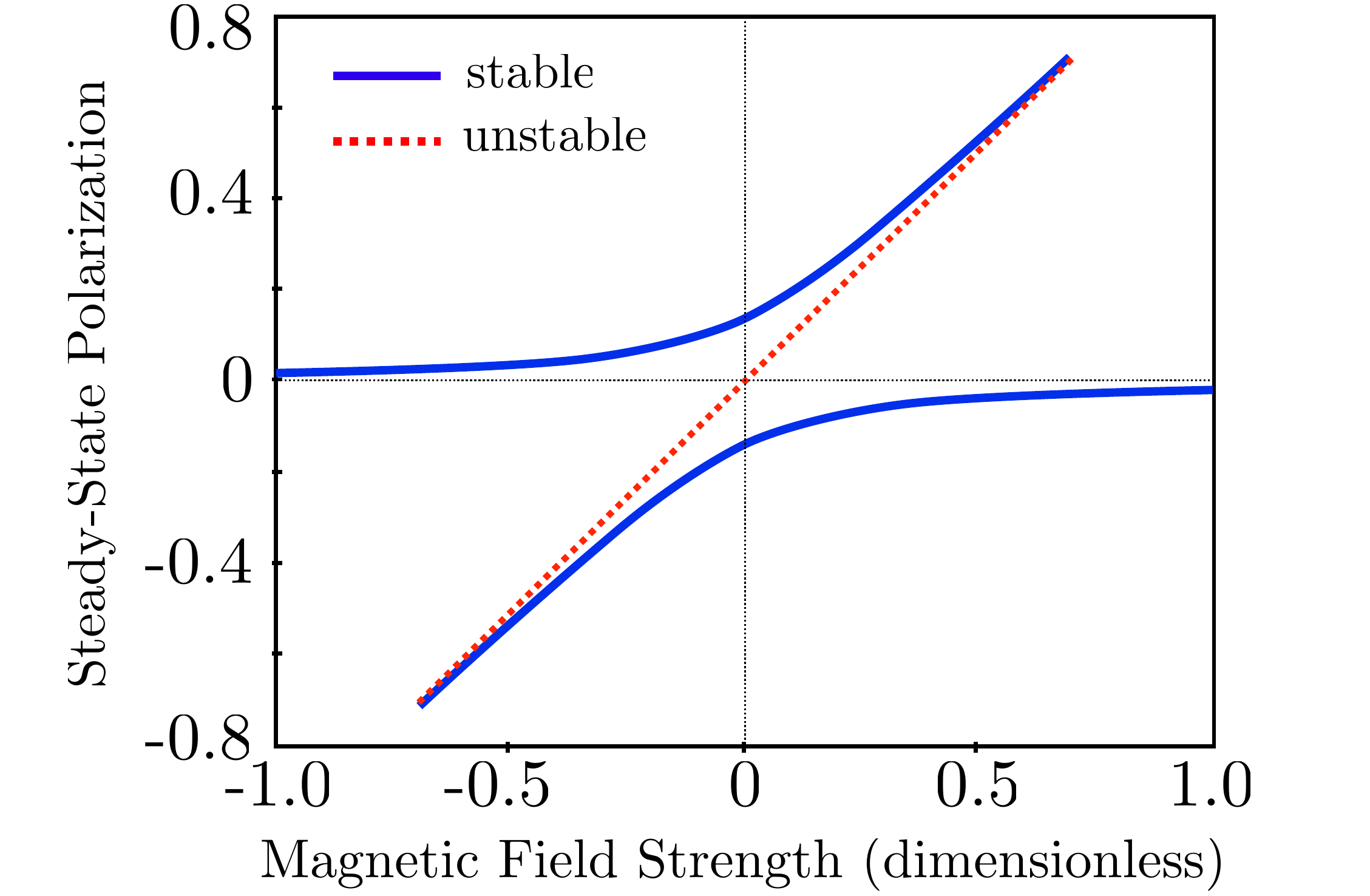}
\caption[]{Polarization fixed points of Eq.(\ref{xdot}), satisfying Eq.(\ref{Fx}), as a function of dimensionless magnetic field strength $\tilde{B}$ for $\tilde\epsilon = \tilde\gamma = 0.01$, $\tilde m = 0$, in absence if spin injection, $\eta = 0$. 
}
\label{fig:FPs}
\end{figure}

A typical pattern of fixed points for systems with $\eta = 0$ is illustrated in the instability diagram in Fig.\ref{fig:FPs}, where we plot the roots of Eq.(\ref{Fx}) as a function of magnetic field $\tilde B$. 
Solid (dotted) lines indicate stable (unstable) fixed points. 
Note that 
in absence of spin injection, i.e for $\eta = 0$, the solutions are symmetric with respect to 
$B$-inversion. The system exhibits bistability over a wide range of magnetic field strengths, with bifurcation points near $|\tilde B| = 0.7$ where bistability disappears.


To investigate the pattern of instabilities in more detail, we examine the 
 discriminant of Eq.(\ref{Fx}), which we denote by $\Delta[F(x)]$. For a general polynomial, the discriminant $\Delta = \prod_{i<j}(x_i - x_j)^2$ is a symmetric function of the polynomial's roots $\{x_i\}$.
Each complex-conjugate pair of roots contributes a factor of -1 to $\Delta$. 
Therefore the bifurcation points, where two real solutions merge and turn into a complex conjugate pair, correspond to the zeros (sign-changing points) of the discriminant.

Because the discriminant $\Delta[F(x)]$ is a symmetric function of the roots of $F(x)$, it can be expressed directly as a polynomial in the coefficients of $F(x)$. For a cubic polynomial of the form (\ref{Fx}), the discriminant is given by\cite{Discriminant}
\be
\label{Discr}\Delta[F(x)] = 18abcd - 4b^3d + b^2c^2 - 4ac^3 - 27a^2d^2.
\ee
Thus the problem of mapping out the bifurcations of the fixed points of the flow $\dot{x}$ in Eq.(\ref{xdot}) is reduced to
the problem of solving for the roots of $\Delta[F(x)]$ in Eq.(\ref{Discr}), with $a, b, c, {\rm and}\ d$ taken from Eq.(\ref{abcd}). 

Because we are primarily interested in the magnetic-field-inversion symmetry/asymmetry of the system, we focus on the 
$B$-dependence of the discriminant $\Delta[F(x)]$. With all other parameters fixed, the equation $\Delta[F(x)] = 0$ yields a fifth-order polynomial in 
$\tilde B$, whose roots determine the bifurcations of the fixed points of the flow (\ref{xdot}).
\begin{figure}[t]
\includegraphics[width=3.4in]{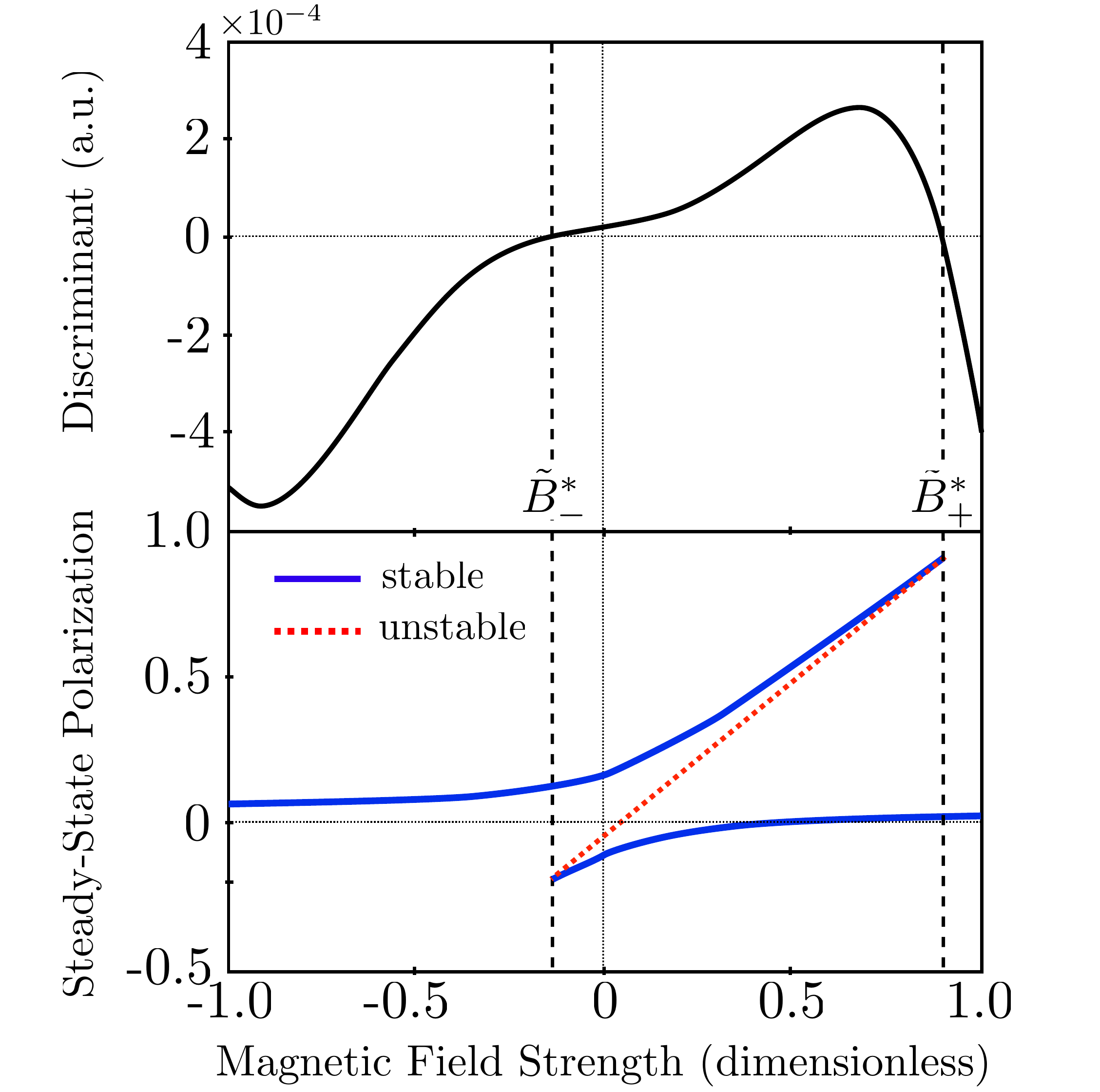}
 \caption[]{Discriminant $\Delta[F(x)]$, Eq.(\ref{Discr}), {\it vs.} dimensionless magnetic field $\tilde B$ for $\tilde\epsilon = 0.01$, $\tilde\gamma = 0.01$, $\eta = 0.05$, and $\tilde m = 0.78$ (upper panel) and the corresponding fixed point diagram (lower panel). The bifurcation points $\tilde B^*_\pm$ at positive and negative field values are indicated by dashed lines. Note that even with small (5\%) spin injection, the instability fields $B^*_+$ and $B^*_-$ differ by one order of magnitude.
}
\label{fig:Discr}
\end{figure}
The full expression for $\Delta[F(x)]$ is quite cumbersome, and we do not reproduce it here.
The expansion of $\Delta$ in the regime $\tilde\epsilon,\tilde\gamma,\tilde m,\eta \ll 1$, 
up to third order in all parameters, reads as
$\Delta \approx \Delta^{(2)} + \Delta^{(3)}$, with
$\Delta^{(2)} = -4(\tilde\gamma^2 + \tilde\epsilon^2)\tilde B^4 + 4\tilde\epsilon^2 \tilde B^2$ and 
$\Delta^{(3)} =  -4\eta (\tilde m\tilde\gamma) \tilde B^5 + 4\eta(\tilde m\tilde\gamma + 4\tilde\gamma^2 + 2\tilde\epsilon^2)\tilde B^3
 - 40\tilde\epsilon(\tilde\epsilon^2 + \tilde\gamma^2)\tilde B^2 -8\eta\tilde\epsilon^2 \tilde B + 32\, \tilde\epsilon^3$.
Note that $\eta$ first appears 
in $\Delta^{(3)}$, which is linear in $\eta$. 
There, $\eta$ multiplies each odd power of $\tilde{B}$, ensuring that the discriminant is invariant under $\eta \rightarrow -\eta, \tilde{B} \rightarrow -\tilde{B}$.
This is a fundamenal symmetry of the system, and holds to all orders.

\begin{figure}[t]
\includegraphics[width=3.4in]{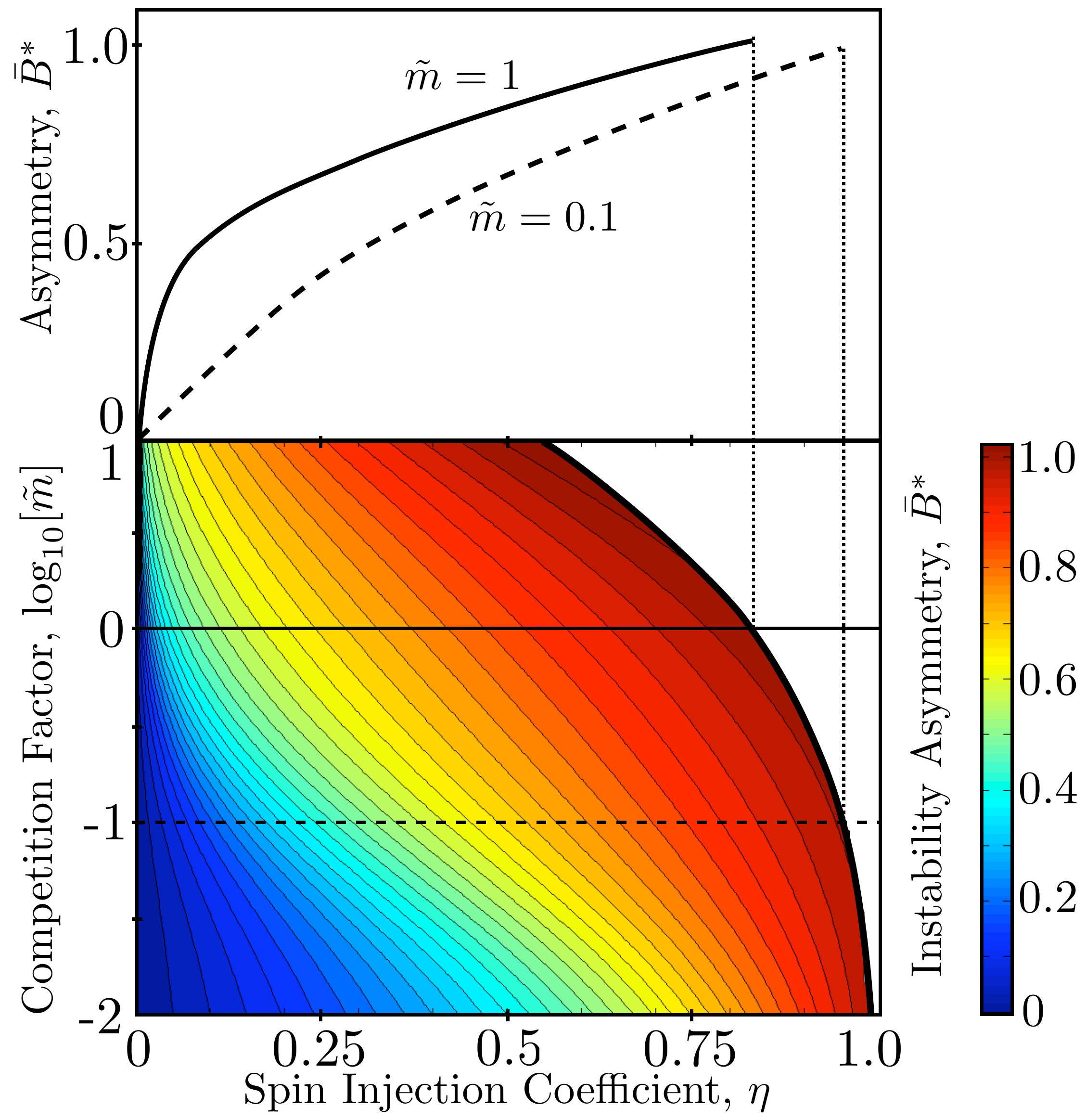}
\caption[]{Dependence of instability asymmetry $\bar{B}^* = \frac12(\tilde{B}^*_+ + \tilde{B}^*_-)$ on the spin-injection coefficient $\eta$ and the factor $\tilde m$, which describes the competition between hyperfine and nuclear-spin-independent decay rates.
}
\label{fig:Asymm2D}
\end{figure}

As demonstrated in Fig.\ref{fig:Discr}, for 
$\eta\neq0$, i.e. when the incident current carries spin-polarization, 
the zeros of $\Delta$ can be highly asymmetric in $\tilde B$. To explore the degree of magnetic-field-inversion symmetry breaking in more detail, in Fig.\ref{fig:Asymm2D} we plot the instability asymmetry parameter $\bar{B}^* = \frac12 (\tilde{B}^*_+ + \tilde{B}^*_-)$ as a function of the spin injection coefficient $\eta$, and the parameter $\tilde{m}$ which describes the competition between the hyperfine transition rates $W_\pm^{\rm HF}$ and the nuclear-spin-independent decay rate $W^{\rm in}$. Here $\tilde B^*_+$ and $\tilde B^*_-$ are the upper and lower bifurcation points, as indicated by the dashed lines in Fig.\ref{fig:Discr}.
For $g^* < 0$, as is typical for GaAs and InAs, two other roots of the fifth-order equation $\Delta=0$ are complex, 
while the fifth root is unphysical
because it corresponds to a 
nuclear spin polarization of greater than 100\%.

For weak spin injection, i.e. for small $\eta$, the instability asymmetry $\bar{B}^*$ grows monotonically with $\eta$. The rate at which $\bar{B}^*$ grows with $\eta$ is controlled by the competition between hyperfine and non-hyperfine decay channels, i.e. by $\tilde m$. As illustrated by the line cuts in the upper panel, the asymmetry grows very sharply with $\eta$
when the two escape processes compete with comparable magnitudes ($\tilde{m} = 1$, solid line). Note that for very large $\eta$, we find a boundary beyond which all bifurcation points disappear, and the system is stable, being partially polarized, for all values of $\tilde B$.


Within the quantum dot, the levels $\Ket{T_+}$ and $\Ket{T_-}$ are defined with respect to a quantization axis which is directed nearly along 
the external field $\bf B$. However, the 
polarization axis of 
electron spin injection is mostly determined by the spin-orbit interaction along the path between the source and the dot. 
In the discussion above, we have implicitly assumed that injected spins were polarized along $\bf B$, in which case  
spin injection directly leads to an imbalance of the probabilities $P_+$ and $P_-$ to load the $\Ket{T_+}$ and $\Ket{T_-}$ states. Suppose instead that the magnetic field is oriented perpendicular to the axis of electron spin injection. In this case, the system will on average have no preference for loading either $\Ket{T_+}$ or $\Ket{T_-}$, and therefore we would find $\eta = 0$. Thus we expect that, within the simplest model of Zeeman-field-independent spin-injection, the spin-injection coefficient $\eta$ should vary like the cosine of the angle between the external field and the spin-injection axis.

In conclusion, spin-orbit coupling results in spin polarization of the electrons injected from nonmagnetic electrodes into a quantum dot even in the absence of an external magnetic field $\bf B$.  
He have shown that $\bf B$-inversion asymmetry of the nuclear polarization instabilities can serve as a highly sensitive tool for detecting this polarization.


We gratefully acknowledge S. M. Frolov and L. P. Kouwenhoven for sharing their data with us, and thank B. I. Halperin for illuminating discussions.
This work was supported by NSF grants DMR-090647 and PHY-0646094 (MR), NSF-MWN (ER) and IARPA (MR,ER). 





\begin{references}

\bibitem{TheBook} L. D. Landau, E. M. Lifshitz, and L. P. Pitaevskii,
{\it Electrodynamics of Continuous Media} (Butterworth Heinemann, Oxford, 1984)

\bibitem{Perel03} V. I. Perel, S. A. Tarasenko, I. N. Yassievich, S. D. Ganichev, V. V. Belkov and W. Prettl, Phys. Rev. B {\bf 67}, 201304(R) (2003).


\bibitem{KiselevKim03} A. A. Kiselev and K. W. Kim, J. Appl. Phys. {\bf 94}, 4001 (2003); Phys. Rev. B {\bf 71}, 153315 (2005).

\bibitem{Ohe} J. Ohe, M. Yamamoto, T. Ohtsuki1, and J. Nitta,  Phys. Rev. B {\bf 72}, 041308(R) (2005).

\bibitem{Eto} M. Eto, T. Hayashi and Y. Kurotani, J. Physi. Soc. Jpn. {\bf 74} (2005).

\bibitem{Krich} J. J. Kirch and B. I. Halperin, Phys. Rev. B {\bf 78}, 035338 (2008).

\bibitem{Sablikov} V. A. Sablikov, Phys. Rev. B {\bf 82}, 115301 (2010).




\bibitem{ZhaiXu} F. Zhai and H. Q. Xu, Phys. Rev. Lett. {\bf 94}, 246601 (2005).

\bibitem{Rokhinson} L. P. Rokhinson, L. N. Pfeiffer, and K. W. West, Phys. Rev. Lett. {\bf 96}, 156602 (2006).

\bibitem{Debray} P. Debray, S. M. S. Rahman, J.Wan, R. S. Newrock, M. Cahay, A. T. Ngo, S. E. Ulloa, S. T. Herbert, M. Muhammad, and M. Johnson, Nat. Nanotechnol. {\bf 4}, 759 (2009).

\bibitem{Ono04} K. Ono and S. Tarucha, Phys. Rev. Lett. {\bf 92}, 256803 (2004).

\bibitem{Koppens05} F. H. L. Koppens, J. A. Folk, J. M. Elzerman, R. Hanson, L. H. Willems van Beveren, I. T. Vink, H. P. Tranitz, W. Wegscheider, L. P. Kouwenhoven, and L. M. K. Vandersypen, Science {\bf 309} 1346 (2005).

\bibitem{Pfund07}
A. Pfund, I. Shorubalko, K. Ensslin, R. Leturcq, Phys. Rev. Lett. {\bf 99}, 036801 (2007).


\bibitem{FK} S. M. Frolov and L. P. Kouwenhoven, private communication.

\bibitem{Rudner1} M. S. Rudner and L. S. Levitov, Phys. Rev. Lett. \textbf{99}, 036602 (2007); Nanotechnology {\bf 21}, 274016 (2010).

\bibitem{Inarrea07} J. I\~{n}arrea, G. Platero, A. H. MacDonald, Phys. Rev. B {\bf 76}, 085329 (2007).

\bibitem{Amasha} S. Amasha, K. MacLean, I. P. Radu, D. M. Zumbuhl, M. A. Kastner, M. P. Hanson, A. C. Gossard, Phys. Rev. B {\bf 78}, 041306(R) (2008) .

\bibitem{Ren10} Y. Ren, W. Yu, S. M. Frolov, J. A. Folk, and W. Wegscheider, Phys. Rev. B {\bf 81}, 125330 (2010).

\bibitem{OnoScience} K. Ono, D. G. Austing, Y. Tokura, and S. Tarucha, Science {\bf 297}, 1313 (2002).

\bibitem{switching} J. Danon, I. T. Vink, F. H. L. Koppens, K. C. Nowack, L. M. K. Vandersypen, and Yu. V. Nazarov, Phys. Rev. Lett. {\bf 103}, 046601 (2009).

\bibitem{Discriminant} {\tt http://en.wikipedia.org/wiki/Cubic\_function}.
\end{references}
\end{document}